\documentclass[aps, prx,twocolumn,longbibliography,superscriptaddress,amsmath,amssymb,floatfix]{revtex4-2}
\usepackage{amssymb}
\usepackage{graphicx}
\usepackage{amsmath}
\usepackage{color}
\usepackage{mathrsfs}
\usepackage{float}
\usepackage{indentfirst}
\usepackage{mathrsfs}
\usepackage{float}
\usepackage{indentfirst}
\usepackage{textcomp}
\usepackage{comment}
\usepackage{mathtools}
\usepackage{natbib,hyperref}
\usepackage{soul}

\begin{document}

\title{Parent Hamiltonian for Fully-augmented Matrix Product States}

\author{Xiangjian Qian}
\affiliation{Key Laboratory of Artificial Structures and Quantum Control (Ministry of Education),  School of Physics and Astronomy, Shanghai Jiao Tong University, Shanghai 200240, China}

\author{Mingpu Qin} \thanks{qinmingpu@sjtu.edu.cn}
\affiliation{Key Laboratory of Artificial Structures and Quantum Control (Ministry of Education),  School of Physics and Astronomy, Shanghai Jiao Tong University, Shanghai 200240, China}

\affiliation{Hefei National Laboratory, Hefei 230088, China}

\date{\today}

\begin{abstract}

Density Matrix Renormalization Group (DMRG) or Matrix Product States (MPS) is the most effective and accurate method for studying one-dimensional quantum many-body systems. But the application of DMRG to two-dimensional systems is not as successful because the limited entanglement encoded in the wave-function ansatz.
Fully-augmented Matrix Product States (FAMPS), introduced recently in Chin. Phys. Lett. 40, 057102 (2023), extends MPS formalism to two dimension and increases the entanglement in the wave-function ansatz, representing a significant advance in the simulation of two-dimensional quantum many-body physics. In the study of one-dimensional systems, the concept of parent Hamiltonian for MPS has proven pivotal in the understanding of quantum entanglement. In this work, we extend this framework to two-dimensional systems. We illustrate the procedure to construct a two-dimensional Hamiltonian with given FAMPS as its exact ground state (the parent Hamiltonian for FAMPS). Additionally, through numerical simulations, we demonstrate the effectiveness of the algorithm outlined in Chin. Phys. Lett. 40, 057102 (2023) in precisely identifying the FAMPS for the constructed parent Hamiltonian. The introduction of FAMPS and its associated parent Hamiltonian provides a useful framework for the future investigations of two-dimensional quantum many-body systems.

\end{abstract}

\pacs{71.27.+a}
\maketitle

{\em Introduction --}
The study of strongly correlated quantum many-body systems is one of the most important themes in condensed matter physics, because exotic quantum
states usually emerges in these systems \cite{RevModPhys.66.763}. However, the accurate solution of strongly correlated quantum many-body systems is usually difficult, necessitating the use of numerical simulations \cite{PhysRevX.5.041041}.  
Density Matrix Renormalization Group (DMRG) or  Matrix Product States (MPS) \cite{PhysRevLett.69.2863,PhysRevB.48.10345,RevModPhys.77.259,SCHOLLWOCK201196,doi:10.1146/annurev-conmatphys-020911-125018,1992CMaPh.144..443F} is a powerful numerical framework for characterizing and simulating one-dimensional (1D) quantum many-body systems. This framework offers a concise but effective representation of 1D quantum states. Previous results have demonstrated the power of DMRG in the study of (quasi) 1D systems \cite{RevModPhys.77.259} and it is now the workhorse for 1D quantum systems. However, the direct application of DMRG to the study of two-dimensional (2D) systems is not as successful as 1D, primarily due to the limited entanglement encoded in the underlying wave-function ansatz \cite{PhysRevB.49.9214}. 

To address this limitation, various ansatzes have been proposed to generalize MPS to 2D, such as Projected Entangled Pair States (PEPS), Multiscale Entanglement Renormalization Ansatz (MERA), and Projected Entangled Simplex States (PESS) \cite{2004cond.mat..7066V,PhysRevLett.99.220405,PhysRevLett.102.180406,PhysRevX.4.011025} and so on \cite{RevModPhys.93.045003,tao_book}. However, these extensions often come with prohibitively high computational costs, making them less popular than DMRG.
  
Fully-augmented Matrix Product States (FAMPS) \cite{Qian_2023} were proposed by considering the trade-off between the entanglement encoded in the ansatz and the overall cost. FAMPS is constructed by augmenting MPS with unitary transformation of physical degrees of freedom, i.e., the so called disentanglers \cite{PhysRevLett.99.220405}. 
It has been demonstrated that FAMPS can support area-law entanglement for 2D systems while maintaining the low cost of DMRG with a small overhead. Numerical results have shown the improvement of accuracy of FAMPS over MPS for a variety of iconic two dimensional spin models \cite{Qian_2023, PhysRevB.109.L161103}. 

It is known that for a given MPS, we can construct the parent Hamiltonian whose ground state is exactly the MPS \cite{1992CMaPh.144..443F,2006quant.ph..8197P}. This framework provides useful tool in the study of 1D quantum many-body systems \cite{RevModPhys.93.045003}, enabling a simple connection between the ground state properties of a quantum system and the local interactions described by its Hamiltonian. However, the parent Hamiltonian for MPS is one-dimensional which limits the application of this framework to two-dimensional quantum many body systems. As FAMPS is a generalization of MPS to two-dimension, it is natural to ask whether we can also construct two dimensional parent Hamiltonian for FAMPS. Such an extension could provide valuable insights in the study of two-dimensional quantum many-body systems.
In the rest of the paper, we will give a positive answer to this question by illustrating the steps to build the parent Hamiltonian for FAMPS.

\begin{figure}[t]
    \includegraphics[width=80mm]{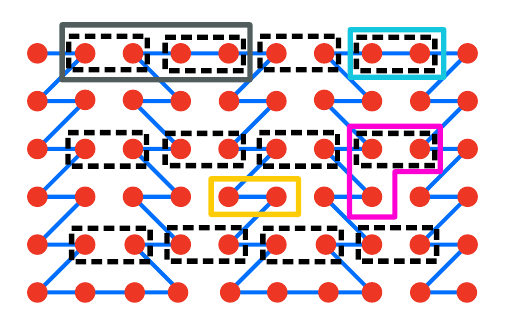}
        \caption{The blue lines show a scheme to arrange an 1D lattice (MPS) into a 2D
        one for the Majumdar-Ghosh model (state) with open boundary conditions. The disentanglers in FAMPS are denoted as dashed black rectangles. The solid rectangles in different colors denote different interaction terms of $H_{\text{FAMPS}}^{\text{MG}}$ in the parent Hamiltonian of FAMPS in Eq.~(\ref{H_FAMPS}). More discussion can be found in the main text.}
        \label{FAMPS_J1J2}
\end{figure}

{\em Parent Hamiltonian for FAMPS --}
DMRG is arguably the workhorse for the study of quasi-one-dimensional quantum many body systems \cite{PhysRevLett.69.2863,PhysRevB.48.10345,RevModPhys.77.259}. The underlying wave-function ansatz of DMRG is MPS \cite{PhysRevLett.75.3537}, which is defined as
\begin{equation}
|\text{MPS} \rangle=\sum_{\{\sigma_i\}} \text{Tr}[A^{\sigma_1}A^{\sigma_2}A^{\sigma_3}\cdots A^{\sigma_n}]|\sigma_1 \sigma_2 \sigma_3\cdots \sigma_n\rangle
\label{MPS}
\end{equation}
where $A$ is a rank-3 tensor with one physical index $\sigma_i$ (with dimension $d$) and two virtual indices (with dimension $D$). 

FAMPS is an extension of MPS by including an extra layer of disentanglers \cite{Qian_2023}, which is connected to the physical indices of MPS. FAMPS is defined as
\begin{equation}
|\text{FAMPS}\rangle=D(u)|\text{MPS}\rangle
\label{FAMPS}
\end{equation}
where $D(u)=\prod_{m} u_m$ denotes the disentangler layer satisfying the unitary condition $D(u)D^\dagger(u) = D^\dagger(u)D(u) = I$ . $u_m$ is the local disentangler which usually involves only two sites \cite{Qian_2023}.  In practical calculations, we need to follow some criteria \cite{Qian_2023} in the construction of FAMPS to make the cost low. Other than that, there are much freedom in the scheme to place disentanglers and the way to rearrange a one dimensional MPS into a two dimensional form. This extension makes FAMPS capable of describing two dimensional systems more accurately than MPS and extends the power of MPS to wider cylinders. At the same time, the (O($D^3$)) low cost of DMRG is maintained with small overhead ({with total cost O($d^4D^3$)}) \cite{Qian_2023}. 

It is known that for each MPS, we can construct a local, frustration-free, 1D parent Hamiltonian with the given MPS being its ground state \cite{1992CMaPh.144..443F,2006quant.ph..8197P}. In the following, we will discuss the construction of parent Hamiltonian for FAMPS defined in Eq.~(\ref{FAMPS}). Suppose we have the parent Hamiltonian for the MPS part in the FAMPS, i.e., $H_\text{MPS}|\text{MPS}\rangle = E_g|\text{MPS}\rangle$. We can easily have  $D(u)H_\text{MPS}D^\dagger(u)D(u)|\text{MPS}\rangle = E_gD(u)|\text{MPS}\rangle$ and $H_\text{FAMPS}|\text{FAMPS}\rangle = E_g|\text{FAMPS}\rangle$ with the definition $H_\text{FAMPS} = D(u)H_\text{MPS}D^\dagger(u)$. So it is obvious that $H_\text{FAMPS}$ is the parent Hamiltonian of $|\text{FAMPS}\rangle$ defined in Eq.~(\ref{FAMPS}). This conclusion is easy to understand because the disentangler-layer $D(u)$ essentially consist of unitary transformations. Because the parent Hamiltonian $H_\text{MPS}$ is local \cite{1992CMaPh.144..443F,2006quant.ph..8197P}, it is obvious that we can also make the interaction local when rearranging the MPS into a 2D form. Disentangler $u_m$ is usually chosen as local unitary transformation, so $H_\text{FAMPS}$ is a local 2D Hamiltonian.


{\em A simple example: the Majumdar-Ghosh state --}
In the following, we take the Majumdar-Ghosh (MG) state \cite{Majumdar_1970} as a concrete example to show the explicit form of $H_\text{FAMPS}$. For the spin-1/2 Heisenberg chain with both nearest ($J_1$) and next nearest ($J_2$) neighboring interactions with the Hamiltonian
\begin{equation}
    H =\sum_{i}J_1S_{i} \cdot S_{i+1}+J_2 S_{i} \cdot S_{i+2},
    \label{H_j1j2}
   \end{equation}
it is known that at the Majumdar-Ghosh point ($J_2/J_1 = 0.5$), the ground state is the product of singlets of all nearest neighboring sites. The ground states have two-fold degeneracy connecting by transnational operation under periodic boundary conditions. The ground state can be represented as an MPS shown in Eq.~(\ref{MPS}) with bond dimension $D = 3$. The local matrices are 
\begin{equation}
    A^{\uparrow}=\begin{bmatrix}  
        0 & 1 & 0\\  
        0 & 0 & 0\\
        \frac{1}{\sqrt{2}} & 0 & 0\\  
    \end{bmatrix},   
    A^{\downarrow}=\begin{bmatrix}  
        0 & 0 & 1\\  
        \frac{-1}{\sqrt{2}} & 0 & 0\\
        0 & 0 & 0\\  
    \end{bmatrix}
\end{equation} 
We first construct a FAMPS based on the MPS representation of the MG state. We only discuss system with open boundary conditions in this work. We can reorganize the 1D MG chain and the corresponding MPS ground state into a 2D form, as illustrated by the blue lines in Fig.~\ref{FAMPS_J1J2}. We then introduce disentanglers, represented by dashed black rectangles in Fig.~\ref{FAMPS_J1J2}, to construct a FAMPS. For simplicity, we only include disentanglers on half the sites of the system. It is known in \cite{Qian_2023} that there is no non-trivial degree of freedom in the two-site disentangler if we want to maintain the SU(2) symmetry. Therefore, in this work, we only consider U(1) symmetry in the disentanglers. As a result, the parent Hamiltonian for FAMPS solely possesses U(1) symmetry. Nevertheless, as shown in \cite{Qian_2023}, we can block two sites into one to maintain the SU(2) symmetry in the disentangler and construct parent Hamiltonian with SU(2) symmetry for FAMPS. For simplicity, we consider the disentanglers with U(1) symmetry as
\begin{equation}
    u_m=\begin{bmatrix}  
        1 & 0 & 0 & 0\\  
        0 &  \frac{1}{\sqrt{2}} &  \frac{1}{\sqrt{2}} & 0\\
        0 & -\frac{1}{\sqrt{2}} & \frac{1}{\sqrt{2}} & 0\\
        0 & 0 & 0 &  1\\   
        \end{bmatrix}
\label{disen}
\end{equation}

\begin{figure}[t]
	\includegraphics[width=90mm]{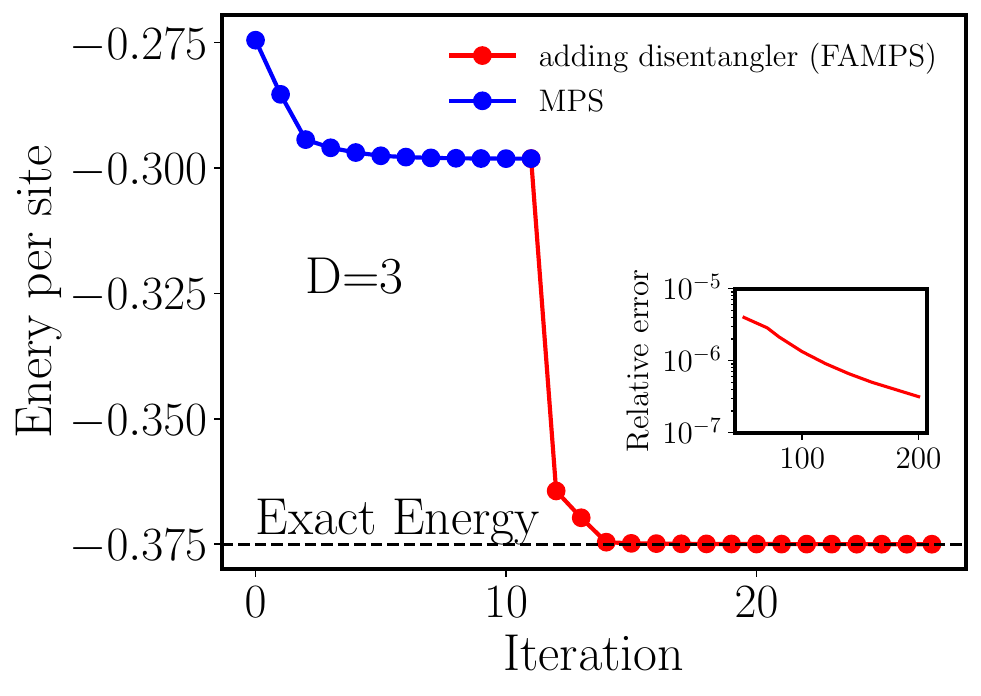}
	\caption{Ground state energy versus optimization step in FAMPS for $H^{\text{MG}}_{\text{FAMPS}}$ in Eq.~(\ref{H_FAMPS}) on a $8\times 8$ lattice with open boundary conditions. The bond dimension is set as $D = 3$. A pure MPS simulation is performed followed by FAMPS simulation with disentanglers. 
		It can be seen that the MPS energy is far away from the exact value, but by including disentanglers, the FAMPS energy converges to the exact value (-0.375) in three steps. The inset shows the relative error of the ground state energy as a function of iteration step in FAMPS.}
	\label{Parent_J1J2_simulation}
\end{figure}

According to the discussion above, the parent Hamiltonian for the MG state induced FAMPS can be obtained as
\begin{equation}
    H_{\text{FAMPS}}^{\text{MG}}=D(u)H_{\text{MPS}}^{\text{MG}}D^\dagger(u)=\prod_{m} u_m H_{\text{MPS}}^{\text{MG}} \prod_{m} u_m^{\dagger}
\end{equation}
with $H_{\text{MPS}}^{\text{MG}}$ the Hamiltonian in Eq.~(\ref{H_j1j2}) with $J_2/J_1 = 0.5$.

Due to the locality of $H_{\text{MPS}}^{\text{MG}}$ (containing only two-site operators) and the unitarity of $u_m$ ($u_mu_m^{\dagger} = u_m^{\dagger}u_m = I$), the interaction terms in $H_{\text{FAMPS}}^{\text{MG}}$ can be classified into four types, illustrated by different colored solid rectangles in Fig.~\ref{FAMPS_J1J2}.
 
For $S_j\cdot S_k$ term with sites $j$ and $k$ connected to two different disentanglers $u_{ij}$ and $u_{kl}$ (${i, j}$ denote the sites involved in the disentangler $u_{ij}$) denoted as solid gray rectangles in Fig.~\ref{FAMPS_J1J2}, we have term in $H_{\text{FAMPS}}^{\text{MG}}$ as
\begin{equation}
    \begin{split}
    h^4_{ijkl}=&\prod_{m} u_m S_j\cdot S_k\prod_{m} u_m^{\dagger}\\=&u_{ij}u_{kl}S_j\cdot S_k u^{\dagger}_{ij}u^{\dagger}_{kl}\\
    =&-\frac{1}{4}\Delta_{ij}\Delta_{kl}+\frac{1}{4}\Delta_{ij}(S_k^z+S_l^z)\\ &
    -\frac{1}{4}\Delta_{kl}(S_i^z+S_j^z)-\Delta_{il}S_j^zS_k^z\\ &
    +\frac{1}{4}(S_i^zS_k^z+S_i^zS_l^z+S_j^zS_k^z+S_j^zS_l^z)\\ &
    +\frac{1}{2}(\Delta_{jl}S_k^z-\Delta_{ik}S_j^z)+\frac{1}{4}\Delta_{jk}
    \end{split}
\end{equation}
where $\Delta_{ij}=S^+_iS^-_j+S^-_iS^+_j$ and $S^+, S^-, S^z$ are Pauli matrices. We can see that $h^4_{ijkl}$ contains interactions involving $4$, $3$ and $2$ sites. 

For $S_i\cdot S_j$ term with only one site ($j$, for example) connected to a disentangler $u_{jk}$, represented by solid pink rectangles in Fig.~\ref{FAMPS_J1J2}, we have term in $H_{\text{FAMPS}}^{\text{MG}}$ as
\begin{equation}
    \begin{split}
    h^3_{ijk}=&\prod_{m} u_m S_i\cdot S_j\prod_{m} u_m^{\dagger}\\=&u_{jk}S_i\cdot S_j u^{\dagger}_{jk}\\
    =&-\frac{1}{2}\Delta_{jk}S_i^z+\frac{1}{2}(S_j^z+S_k^z)S_i^z\\ &+\frac{\sqrt{2}}{4}\Delta_{ij}+\frac{\sqrt{2}}{2}\Delta_{ik}S_j^z
    \end{split}
\end{equation}
We can see that $h^3_{ijk}$ contains interactions involving $3$ and $2$ sites. 

For $S_i\cdot S_j$ term that sites $i$ and $j$ are connected by a single disentangler $u_{ij}$ (solid light blue rectangles in 
Fig.~\ref{FAMPS_J1J2}), we have the term in $H_{\text{FAMPS}}^{\text{MG}}$ as
\begin{equation}
    \begin{split}
    h^2_{ij}=&\prod_{m} u_m S_i\cdot S_j\prod_{m} u_m^{\dagger}\\=&u_{ij}S_i\cdot S_j u^{\dagger}_{ij}\\
    =&\frac{1}{2}(S_i^z-S_j^z)+S_i^zS_j^z
    \end{split}
\end{equation}
We can see that the original Heisenberg term $S_i\cdot S_j$ is transformed into an Ising-type interaction with the disentangler defined in Eq.~(\ref{disen}). It is noteworthy that the ground state of $S_i\cdot S_j$ term is a singlet, representing a maximally entangled state, while the ground state of $h^2_{ij}$ is a product state. This transformation provides a vivid example on the usefulness of disentangler to reduce the entanglement in a quantum system.

\begin{figure}[t]
	\includegraphics[width=80mm]{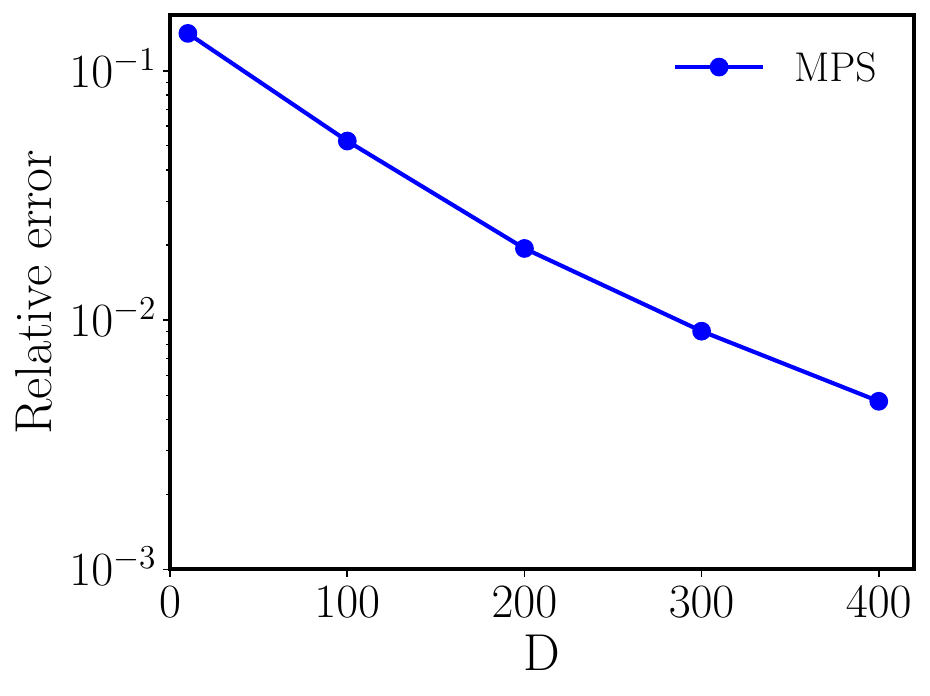}
	\caption{Relative error of the ground state energy as a function of bond-dimension $D$ in pure MPS (DMRG) simulation for the same system in Fig.~\ref{Parent_J1J2_simulation}. It can be seen that the relative error of energy in MPS is still at the level of $10^{-3}$ even with bond dimension $D = 400$, while FAMPS can basically give the exact energy with $D = 3$ as shown in Fig.~\ref{Parent_J1J2_simulation}.}
	\label{Parent_J1J2_simulation_MPS}
\end{figure}

\begin{figure*}[t]
	\includegraphics[width=80mm]{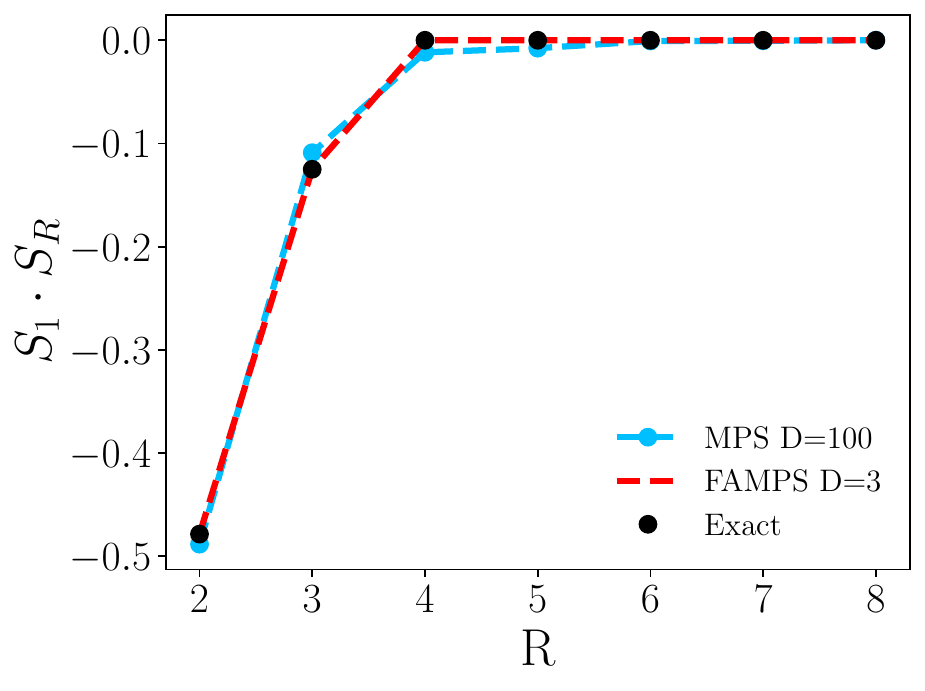}
        \includegraphics[width=80mm]{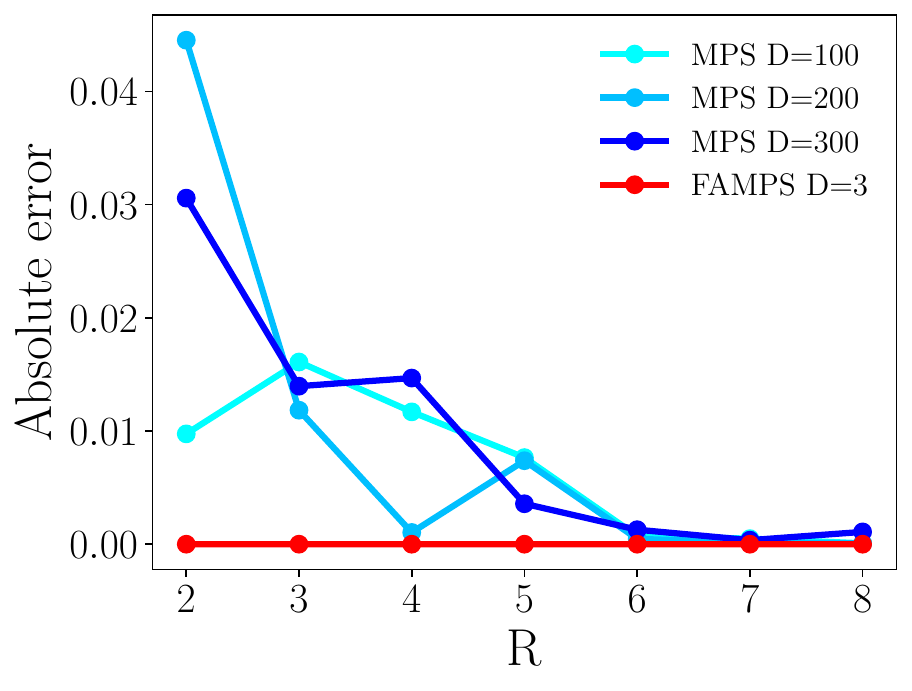}
	\caption{{The spin correlation function along one of the horizontal lines where disentanglers are placed (see Fig.~\ref{FAMPS_J1J2}). Results from FAMPS, MPS and exact calculation are shown. The left panel shows the spin correlation $\langle S_1 \cdot S_R \rangle$ as a function of distance $R$. The right panel shows the absolute error of the MPS and FAMPS results with respect to the exact values. We can see that same as the energy, FAMPS with $D=3$ gives the exact answer, while MPS results with $D=100$ have considerable errors. }}
	\label{CORR}
\end{figure*}
 
For $S_i\cdot S_j$ term not connected to any disentangler (solid yellow rectangles in Fig.~\ref{FAMPS_J1J2}), the term in $H_{\text{FAMPS}}^{\text{MG}}$ is unchanged because
\begin{equation} 
    \begin{split}
    h^{1}_{ij}=&\prod_{m} u_m S_i\cdot S_j\prod_{m} u_m^{\dagger}\\=&S_i\cdot S_j
    \end{split}
\end{equation}

To conclude, $H_{\text{FAMPS}}^{\text{MG}}$ contains interactions involving $4$, $3$, and $2$ sites and also onsite terms. Formally, $H_{\text{FAMPS}}^{\text{MG}}$ can be written as
\begin{equation}
H_{\text{FAMPS}}^{\text{MG}}=\sum_{i}J_ih^4_i+\sum_{j}J_jh^3_j+\sum_{k}J_kh^2_k+\sum_{l}J_lh^{1}_l
\label{H_FAMPS}
\end{equation}
where $h^i$ denotes interaction term involving $i$ sites with interaction strength $J_i$.

We also study the Hamiltonian in Eq.~(\ref{H_FAMPS}) on a $8 \times 8$ square lattice with open boundary conditions, using both FAMPS and MPS. First we perform a FAMPS calculation for the Hamiltonian in Eq.~(\ref{H_FAMPS}). The results are shown in Fig.~\ref{Parent_J1J2_simulation}. In the calculation, we first perform a FAMPS simulation without disentanglers (basically a pure MPS calculation). Then we include disentanglers (initialized as identity) upon the MPS, and perform FAMPS simulations by optimizing the MPS and disentanglers iteratively, following the algorithm in \cite{Qian_2023}. As discussed above, the bond dimension of the exact MPS representation of the MG state is $D = 3$ which means FAMPS with $D = 3$ is enough for the exact ground state of $H_{\text{FAMPS}}^{\text{MG}}$. So we set $D = 3$ in the FAMPS calculation. As shown in Fig.~\ref{Parent_J1J2_simulation}, the MPS energy with $D = 3$ is far away from the exact ground state energy of $H_{\text{FAMPS}}^{\text{MG}}$ ($E_g = -0.375$). However, 
after adding disentanglers, we can find that the energy converges to the exact value in $3$ steps. In the inset of Fig.~\ref{Parent_J1J2_simulation}, we show the relative error of FAMPS energies versus the optimization step in FAMPS, from which we can see the convergence of the energy to the exact value. This result aligns well with our expectations and validates the algorithm proposed in Ref~\cite{Qian_2023}.

We also perform a pure MPS (DMRG) simulation with different bond dimensions for the same system. The results are shown in Fig.~\ref{Parent_J1J2_simulation_MPS}. We can see that the relative error of energy in MPS is still at the level of $10^{-3}$ even with bond dimension $D = 400$, while FAMPS can basically give the exact energy with $D = 3$.
{

We also calculate the spin correlation for the ground state of the resulting parent FAMPS Hamiltonian $H_{\text{FAMPS}}^{\text{MG}}$.
For the original Majumdar-Ghosh model, only the nearest neighboring spin correlation  $\langle S_1 \cdot S_2 \rangle$ is nonzero. After applying the disentanglers, the correlation extends horizontally as shown in the left panel of Fig.~\ref{CORR}, causing the model to begin exhibiting 2D characteristics. The right panel of Fig.~\ref{CORR} also demonstrates that MPS struggles to faithfully represent the ground state of the resulting FAMPS parent Hamiltonian, while FAMPS can precisely represent the ground state with only D=3, similar as we find in the energy results.

}

{
{\em Other examples --}
It can be easily shown that the parent Hamiltonian for the cluster state on the two dimensional Lieb lattice \cite{PhysRevB.109.195420} can be transformed to decoupled chains by special disentanglers (details in the supplementary materials). If we consider the reversal process, by applying the conjugate of the disentanglers to the ground states of the decoupled chains, which are essential MPS \cite{RevModPhys.93.045003}, we can obtain a 2D state with a 2D parent Hamiltonian. Similar constructions can also be found in \cite{PhysRevA.82.052309}, where decoupled Affleck-Kennedy-Lieb-Tasaki (AKLT) state in 1D is transformed by disentanglers to a 2D state with a 2D parent Hamiltonian.       
}

{\em Discussion --} The parent Hamiltonian for any given FAMPS can be constructed following the Majumdar-Ghosh case. First we can construct the parent Hamiltonian $H_\text{MPS}$ for the MPS part in FAMPS, then we can construct the parent Hamiltonian for FAMPS $H_\text{FAMPS}$ by transforming each term in $H_\text{MPS}$ with the disentanglers in FAMPS. {We want to emphasize that even though the parent Hamiltonian for the MPS part in FAMPS is one-dimensional, the resulting parent Hamiltonian is two-dimensional because the disentanglers are non-local in the 1D setup as illustrated in Fig.~\ref{FAMPS_J1J2} (in the 1D setup, the disentanglers span sites with distance $l$ for a $l \times l$ lattice).} The form of the parent Hamiltonian for FAMPS can be complicated depending on the specific structure of FAMPS. {But $H_\text{FAMPS}$ is essentially a local 2D Hamiltonian if the disentanglers are local in 2D}. It will be interesting to explore the variation in FAMPS, i.e., generalizing the disentangler layer $D(u)$ to other unitary transformations, to see whether we can construct $H_\text{FAMPS}$ in the simplest form, i.e., with only nearest neighboring interactions in 2D. {We also note that FAMPS can belong to a phase different from the underlying MPS, because the disentanglers (unitary transformations) are non-local and the constraint of symmetry on disentanglers can be released. An example is the cluster state. It is a symmetry protected topological state which can be transformed from the product state by disentanglers.}

{\em Conclusions --}
FAMPS \cite{Qian_2023} was proposed to alleviate the difficulty of DMRG in the study of two-dimensional quantum many-body systems.
In this work, we establish that FAMPS can serve as the exact ground state for certain two-dimensional Hamiltonians. Utilizing the Majumdar-Ghosh state as an illustrative example, we outline the procedure to construct the parent Hamiltonian for FAMPS. Through numerical simulations, we show that the algorithm proposed in Ref~\cite{Qian_2023} can easily find the exact FAMPS for its parent Hamiltonian. The existence of 2D parent Hamiltonian for FAMPS aligns with the conclusion in Ref~\cite{Qian_2023} that FAMPS can be used to efficiently simulate 2D quantum many body systems.
The construction of parent Hamiltonians for FAMPS extends the framework of parent Hamiltonian for MPS in one-dimensional systems to high dimensional systems, providing a useful framework for the future study of high dimensional quantum many body systems. We can also construct uncle Hamiltonian for FAMPS based on the so-called uncle Hamiltonian for MPS \cite{2015CMaPh.333..299F}. It will be interesting to consider the parent Hamiltonian for FAMPS with fermionic degrees of freedom in the future. 

\begin{acknowledgments}
The calculation in this work is carried out with TensorKit \footnote{The code is developed with TensorKit package at https://github.com/Jutho/TensorKit.jl}. The computation in this paper were run on the Siyuan-1 cluster supported by the Center for High Performance Computing at Shanghai Jiao Tong University. MQ acknowledges the support from the National Key Research and Development Program of MOST of China (2022YFA1405400), the Innovation Program for Quantum Science and Technology (2021ZD0301902),
the National Natural Science Foundation of China (Grant
No. 12274290) and the sponsorship from Yangyang Development Fund.

\appendix

\section{Cluster states}
Let $|0\rangle$ and $|1\rangle$ be the eigenstates of $\sigma^z$. The Cluster state is described by a graph $G=(V,E)$ with the sets vertices of $V$ and edges $E$, where the elements of $V$ and $E$ correspond to qubits initialized in the  eigenstate of $\sigma^x$ with eigen value $1$, i.e., $|+\rangle = (|0\rangle+|1\rangle)/\sqrt{2}$ and Controlled-Z  gates ($CZ = (1 + \sigma^z_i + \sigma^z_{i+1} - \sigma^z_i\sigma^z_{i+1})/2$), respectively:

\begin{equation} 
    |\psi\rangle=\prod_{ij\in E} CZ_{ij} \bigotimes_{i \in V} |+\rangle
\label{cluster_def}
\end{equation}

The state $|\psi\rangle$  is the unique ground state of the stabilizer Hamiltonian $H=-\sum_{i\in V} \sigma^x_i \prod_{j|(ij)\in E} \sigma^z_j $ \cite{PhysRevLett.129.090501}.

It is already known that when the graph G forms a one-dimensional chain, the cluster state $|\psi\rangle$ can be represented by an MPS with $D=2$ with the parent Hamiltonian $H=-\sum_i \sigma^z_i\sigma^x_{i+1}\sigma^z_{i+2}$ \cite{RevModPhys.93.045003}.

In contrast, cluster states on two dimension can be efficiently represented using FAMPS.  By treating the $CZ$ gates as disentanglers, the cluster state can be efficiently presented by FAMPS with the parent Hamiltonian $H=-\sum_{i\in V} \sigma^x_i \prod_{j|(ij)\in E} \sigma^z_j$. 

\begin{figure}[t]
	\includegraphics[width=80mm]{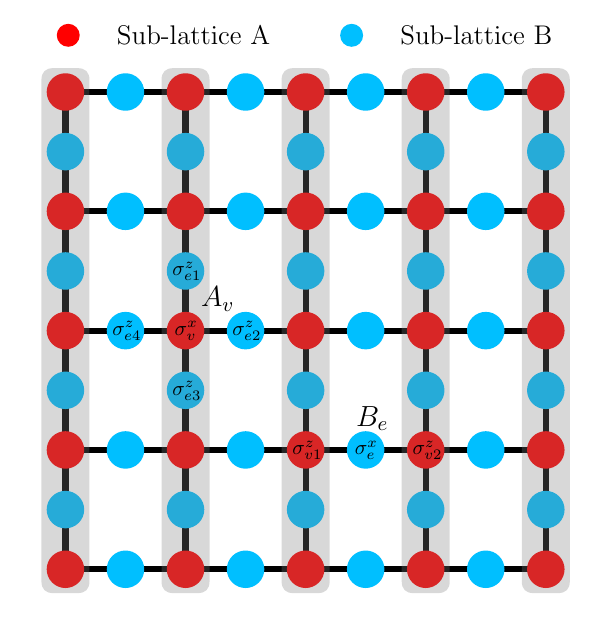}
	\caption{Schematic representation of the 2D cluster state on a Lieb lattice. The Hamiltonian is shown in Eq.~\ref{cluster_H} which contains two types of interactions denoted by $A_v$ and $B_e$. Columns C and D in Eq.~\ref{Cluster_2} are represented by shadowed and non-shadowed columns, respectively. }
	\label{Cluster}
\end{figure}

An example is shown in Fig.~\ref{Cluster}, proposed in Ref.~\cite{PhysRevB.109.195420}. Here, we consider a 2D Lieb lattice, which has two inequivalent sub-lattices. Sub-lattice A is defined on
the vertex of an $N \times N$ square lattice and sub-lattice B is defined at the edges of the square lattice. We put a spin-1/2 on
each of A and B, which is labeled by $v$ and $e$, respectively;
see Fig.~\ref{Cluster}. A 2D cluster state (defined in Eq.~(\ref{cluster_def})) is given by the ground state
of the following frustration-free stabilizer Hamiltonian:

\begin{equation} 
    H_{\text{Cluster}}=-\sum_{v\in A}A_v-\sum_{e \in B}B_e
    \label{cluster_H}
\end{equation}

where $A_v=\sigma^x_v\sigma^z_{e1}\sigma^z_{e2}\sigma^z_{e3}\sigma^z_{e4}$ and $B_e=\sigma^x_e\sigma^z_{v1}\sigma^z_{v2}$.
By introducing disentanglers ($CZ^{\dagger} = CZ$) in each row of the $N\times N$ square lattice, the Hamiltonian $H_{\text{Cluster}}$ is disentangled to a one-dimensional cluster Hamiltonian decorated with some single $\sigma^x$ terms whose ground state can be efficiently represented by MPS:

\begin{equation} 
    H_{\text{Cluster}}^{\prime}=-\sum_{i\in C} \sigma^z_i\sigma^x_{i+1}\sigma^z_{i+2}-\sum_{i \in D}\sigma^x_i
    \label{Cluster_2}
\end{equation}

where $C, D$ represents different columns of the Lieb lattice as shown in Fig.~\ref{Cluster}. Thus, the ground state of $H_{\text{Cluster}}$ can be efficiently represented by FAMPS. 

\end{acknowledgments}

\bibliography{FAMPS-Parent}

\end{document}